\documentclass{PoS}

\title{The ALICE trigger system for LHC Run 3}

\ShortTitle{ALICE Run 3 trigger}

\author{\speaker{M. Krivda}
, D. Evans, K.L. Graham, A. Jusko, R. Lietava, O. Villalobos Baillie and N. Zardoshti\\
        School of Physics and Astronomy, The University of Birmingham, Edgbaston, Birmingham, UK B15 2TT\\
        E-mail: \email{marian.krivda@cern.ch}}

\author{M. Bombara and M. {\v S}ef{\v c}{\'i}k\\
        Faculty of Science, P.J. {\v S}af{\'a}rik University, Ko{\v s}ice, Slovakia\\ }

\author{I Kr{\' a}lik\\
        Institute of Experimental Physics, Slovak Academy of Sciences, Ko{\v s}ice, Slovakia\\ }

\author{L.A. P{\'e}rez Moreno \\
        Benem{\'e}rita Universidad Aut{\'o}noma de Puebla, Puebla, Mexico\\ }

\author{for the ALICE Collaboration.\\ \\}

\abstract{The ALICE Central Trigger Processor (CTP) is going to be upgraded for LHC Run 3 with completely new hardware and a new Trigger and Timing Control (TTC-PON) system based on a Passive Optical Network (PON) system. The new trigger system has been designed as dead time free and able to transmit trigger data at 9.6 Gbps. A new universal trigger board has been designed, where by changing the FMC card, it can function as a CTP or as a LTU. It is based on the Xilinx Kintex Ultrascale FPGA and upgraded TTC-PON. The new trigger system and the prototype of the trigger board will be presented.}

\FullConference{Topical Workshop on Electronics for Particle Physics\\
		11 - 14 September 2017\\
		Santa Cruz, California, U.S.A.}

\begin{document}

\section{Introduction}

In run 3 of the LHC, it is envisaged that the luminosity for Pb-Pb collisions will increase significantly, leading to a 50 kHz interaction rate for Pb-Pb, while in pp and pA mode the interaction rate at Point 2 can be kept to 200 kHz. In these conditions the current readout system requires a major upgrade. The principal tracking detector in ALICE is a Time Projection Chamber (TPC) with a maximum drift time of about 100 $\mu$s. At 50 kHz pile-up is inevitable, presenting a serious challenge owing to the very high multiplicity of Pb-Pb events. The physics signals of interest to ALICE are in general complex to separate, and therefore not amenable to hardware triggers. Instead, ALICE intends to read out {\sl continuously}, applying a minimum bias trigger to the data stream to flag events and using sophisticated filters on fully reconstructed events. In order to do this, most but not all detectors will upgrade to continuous readout. Adapting to continuous readout necessitates a new trigger system, but, as not all the detectors will be upgraded, the new system must retain backwards compatibility, making the system more complicated.
\section{Description of System}

\begin{figure}[t]
\begin{center}

\includegraphics[width=0.75\textwidth]{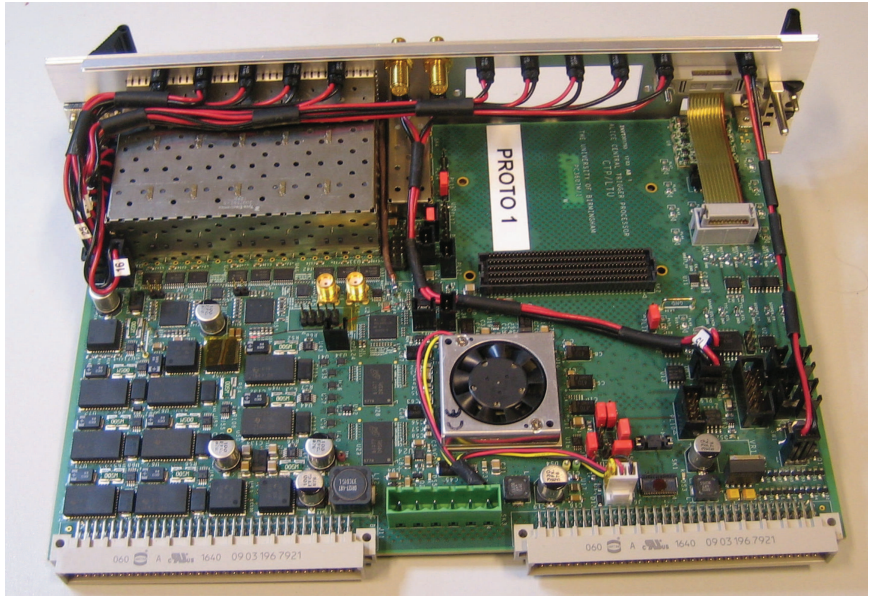}

\end{center}
\caption{ Top view of the prototype board. }
\label{fig-photo}
\end{figure}

Owing to major advances in the processing power of FPGAs since the time when the original CTP was designed, the functionality of the six-board run 1 trigger system can now be accommodated on a single triple-width 6U VME board. The VME format is used only for power supply, with the board using +5V/10A and $\pm12$V/1A, and any other  required voltages being generated on the board using DC-DC converters. The prototype board is shown in figure \ref{fig-photo}. The board is equipped with a Xilinx Kintex Ultrascale FPGA (XCKU040FFVA1156) \cite{fpga}, two 1GB DDR4 memories \cite{mem}, two Si5345 PLLs \cite{pll}, an FMC-HPC connector, two sixfold SFP+ cages and one single SFP+ cage, and two UCD90120A power controllers.

An important feature of the design is that all the data interfaces for the board are via the front panel, rather than via the VME bus. The available interfaces are:

\begin{enumerate}
\item USB-JTAG (for JTAG access to the FPGA and FMC card);
\item IPbus (control and monitoring of the board);
\item DDR4 (access to snapshot memory and trigger data);
\item TTC-PON (new clock and trigger distribution based on Passive Optical Network (PON) system);
\item TTC (legacy RD12 (Run1) Trigger, Timing and Control (TTC) system);
\item GBT (for clock and trigger distribution directly to detector FEE);
\item I2C;
\item SPI;
\item Power Management (PMbus) connector.
\end{enumerate}

In the original (Run 1/Run 2) trigger system, the CTP communicates with each detector via a 6U VME Local Trigger Unit (LTU), in order to provide a uniform interface between the CTP and each detector. This approach will be maintained in Run 3. In the new system  there is a universal trigger board that can function either as a CTP or as one of several types of LTU (TTC-PON, GBT, RD-12 TTC interfaces) with different firmware versions according to the detector FE requirements.

An important feature of the board in LTU mode is the ability to function in standalone mode as a trigger emulator, producing trigger sequences that obey the same protocol as those generated by the physics triggers.
\section{Status of Test Board}

\begin{figure}[t]
\begin{center}

\includegraphics[width=0.75\textwidth]{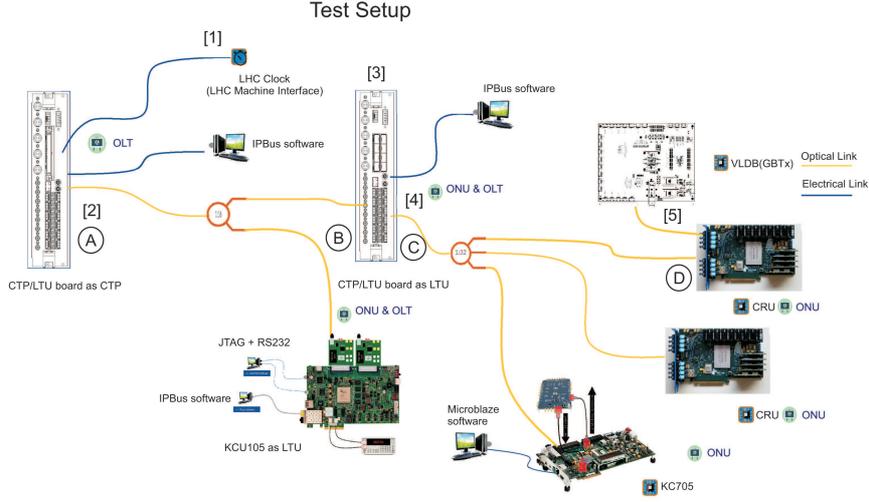}

\end{center}
\caption{ Setup used to evaluate the new trigger system. }
\label{fig-test}
\end{figure}

A prototype board for the new design was delivered in March 2017 and has been extensively tested. The tests reflect both the board itself and the new system of communication of the board with the detectors and the readout chain, which replaced (for the most part) the RD-12 TTC system used up till now for transmission of ALICE trigger signals. In addition to essential tests on voltage stability, temperature stability, FPGA operation and flash memory integrity, more detailed performance using the setup shown in figure \ref{fig-test}. This allowed tests of BER and jitter in the system to be performed.

The TTC-PON system is bidirectional, but the transmission rates in each direction are not the same. In the downstream direction it is 9.6 Gbps, while upstream it is only 2.4 Gbps. No errors were recorded during the test, and  a limit of BER $< 10^{-15}$ in the downstream direction (BER $< 10^{-14}$ in the upstream direction) were achieved. The different limits simply reflect the different transmission rates and the length of time available for the tests.

Jitter measurements were performed at several points in the system, separating the contributions into random, deterministic and periodic components according to the prescription given in \cite{jitter}. (See table \ref{Tab:jitter}.) The LHC clock, as received from the RF2TTC interface, has an overall jitter of about 40 ps. Although the clock degenerates as the signals are transmitted through the system, the use of the Si5345 jitter cleaner at suitable points in the chain means that clock quality can be recovered, giving a very satisfactory performance overall. Somewhat better results are obtained if the very high quality clock from the CG635 generator is substituted for the LHC clock.

\begin{table} [b]
\caption{Jitter Measurements for the new trigger system. The columns indicate: (T$_{j}$) overall jitter for BER~$< 10^{-12}$, (R$_{j}$) random jitter, (D$_{j}$) deterministic jitter and (P$_{j}$) periodic jitter. The rows indicate the places where the jitter was measured. For the first set of measurements, the positions where the jitter was measured are indicated by the numbers in square brackets in the table and in figure \ref{fig-test}. For the second set of measurements, the LHC clock was replaced by a CG635 clock generator.}
\label{Tab:jitter}
\begin{center}
\small

\begin{tabular}{lrrrr}
\hline
  &  T$_{j}$ (ps) & R$_{j}$ (ps)  &  D$_{j}$ (ps) & P$_{j}$ (ps)\\ \hline
RF2TTC local osc. [1]  &  149.2   & 2.24 & 117.54  & 72.04          \\
CTP Si5345 out [2] & 62.14 & 1.21 & 44.94 & 21.88     \\
LTU recovered clock from ONU [3] & 438.36 & 25.88 & 69.23 & 137.44 \\
LTU Si5435 out [4]   & 42.10 & 1.64  & 18.78  & 19.75  \\
GBTx elink0 [5] & 298.95 & 13.00 & 113.53 & 101.91 \\
\hline \hline
CG635  & 110.02 & 7.72  & 0.09 & 7.40 \\
CTP Si5435 out & 17.21  & 1.20 & 0.16 & 1.93 \\
LTU recovered clock from ONU & 340.09 & 17.62 & 88.65 & 134.52 \\
LTU Si5345 out & 17.42 & 1.21 & 0.18 & 2.54 \\
GBTx elink0 & 144.99 & 9.21 & 13.67 & 31.95\\
\hline
\end{tabular}
\end{center}
\end{table}
\section{Summary}
The major upgrade of the ALICE detector in LHC run 3, and the different readout strategy to be employed by the majority of the detectors, means that a new trigger system is necessary. The principal features of the new system, comprising a Central Trigger Processor, a  Local Trigger Unit and a new transmission system (TTC-PON) between the components. A prototype board has been produced and has been extensively tested. Measurements of BER and jitter for the new TTC-PON chain show excellent performance for this new distribution system.

\end{document}